\documentclass[fleqn]{2020SCGE}
\usepackage{color}
\usepackage{float}
\usepackage{gensymb}
\usepackage{url}
\usepackage{hyperref}
\usepackage{booktabs} 
\usepackage{threeparttable}
\usepackage{multirow} 
\usepackage{tabularx}
\usepackage{makecell}
\usepackage{url}     

\DeclareOldFontCommand{\bf}{\normalfont\bfseries}{\mathbf} 
\providecommand{\DIFdeltex}[1]{} 
\RequirePackage{listings} 
\lstdefinelanguage{DIFcode}{ 
  moredelim=[il][\color{white}\tiny]{\%DIF\ <\ }, 
  moredelim=[il][\sffamily\bfseries]{\%DIF\ >\ } 
} 
\lstdefinestyle{DIFverbatimstyle}{ 
	language=DIFcode, 
	basicstyle=\ttfamily, 
	columns=fullflexible, 
	keepspaces=true 
} 
\lstnewenvironment{DIFverbatim}{\lstset{style=DIFverbatimstyle}}{} 
\lstnewenvironment{DIFverbatim*}{\lstset{style=DIFverbatimstyle,showspaces=true}}{} 
\begin{document}

\ensubject{subject}


\ArticleType{Article}
\SpecialTopic{}
\Year{2026}
\Month{October}
\Vol{69}
\No{10}
\DOI{10.1007/s11433-026-3030-x}
\ArtNo{109511}
\ReceiveDate{March 27, 2026}
\AcceptDate{May 27, 2026}
\OnlineDate{August 27, 2026}

\title{Fast radio bursts, magnetars and earthquakes: their ``family feud''?}{Fast radio bursts, magnetars and earthquakes: their ``family feud''?}

\author[1,2]{\\Si-Lu Xu}{}
\author[1]{Yong-Kun Zhang}{ykzhang@nao.cas.cn}
\author[1,3]{Pei Wang}{wangpei@nao.cas.cn}
\author[4,1]{Di Li}{dili@tsinghua.edu.cn}
\author[1,2]{Jun-Shuo Zhang}{}
\author[5]{Tian-Cheng Lv}{}
\author[5,6]{\\Yong-Feng Huang}{}
\author[3,7]{Tian-Cong Wang}{}
\author[8]{Long-Xuan Zhang}{}
\author[8]{Pei-Xin Zhu}{}
\author[1,2]{Jin-Huang Cao}{}
\author[9,10]{\\Yi Feng}{}
\author[3,7]{He Gao}{}
\author[11]{Jian Li}{}
\author[1,2]{Wan-Jin Lu}{}
\author[12,13]{Chen-Chen Miao}{}
\author[14]{Chen-Hui Niu}{}
\author[1,2]{\\Qing-Yue Qu}{}
\author[1,2,3]{Chao-Wei Tsai}{}
\author[1,2]{Yi-Dan Wang}{}
\author[1,2]{Wen-Ting Wang}{}
\author[15,16]{Su-Ming Weng}{}
\author[17]{\\Jia-Fu Wu}{}
\author[18]{Ru-Shuang Zhao}{}
\author[8,19]{Yuan-Chuan Zou}{}
\author[1,2]{Yu-Hao Zhu}{}
\author[17]{Ya-Biao Wang}{}

\footnote{*Corresponding author:(Yong-Kun Zhang, email: ykzhang@nao.cas.cn; Pei Wang, email: wangpei@nao.cas.cn; Di Li, email: dili@tsinghua.edu.cn)}


\AuthorMark{S.-L. Xu}

\AuthorCitation{S.-L. Xu, Y.-K. Zhang, P. Wang, D. Li, J.-S. Zhang, T.-C. Lv, Y.-F. Huang, T.-C. Wang, L.-X. Zhang, P.-X. Zhu, J.-H. Cao, Y. Feng, H. Gao, J. Li, W.-J. Lu, C.-C. Miao, C.-H. Niu, Q.-Y. Qu, C.-W. Tsai, Y.-D. Wang, W.-T. Wang, S.-M. Weng, J.-F. Wu, R.-S. Zhao, Y.-C. Zou, Y.-H. Zhu, and Y.-B. Wang}

\address[{\rm1}]{State Key Laboratory of Radio Astronomy and Technology, National Astronomical Observatories, Chinese Academy of Sciences, Beijing {\rm 100101}, China}
\address[{\rm2}]{University of Chinese Academy of Sciences, Beijing {\rm 100049}, China}     
\address[{\rm3}]{Institute for Frontiers in Astronomy and Astrophysics, Beijing Normal University, Beijing {\rm 102206}, China}
\address[{\rm4}]{New Cornerstone Science Laboratory, Department of Astronomy, Tsinghua University, Beijing {\rm 100084}, China}
\address[{\rm5}]{School of Astronomy and Space Science, Nanjing University, Nanjing {\rm 210023}, China}
\address[{\rm6}]{Key Laboratory of Modern Astronomy and Astrophysics, Ministry of Education (Nanjing University), Nanjing {\rm 210023}, China}
\address[{\rm7}]{School of Physics and Astronomy, Beijing Normal University, Beijing {\rm 100875}, China}
\address[{\rm8}]{School of Physics, Huazhong University of Science and Technology, Wuhan, {\rm 430074}, China}
\address[{\rm9}]{Research Center for Astronomical Computing, Zhejiang Laboratory, Hangzhou {\rm 311100}, China}
\address[{\rm10}]{Institute for Astronomy, School of Physics, Zhejiang University, Hangzhou {\rm 310027}, China}
\address[{\rm11}]{Department of Astronomy, School of Physical Sciences, University of Science and Technology of China, Hefei {\rm 230026}, China.}
\address[{\rm12}]{Department of Astronomy, College of Physics and Electronic Engineering, Qilu Normal University, Jinan {\rm 250200}, China}
\address[{\rm13}]{Shandong Key Laboratory of Space Environment and Exploration Technology, No. 180, Wenhua West Road, Weihai City, Shandong Province {\rm 264209}, China}
\address[{\rm14}]{Institute of Astrophysics, Central China Normal University, Wuhan {\rm 430079}, China}
\address[{\rm15}]{State Key Laboratory of Dark Matter Physics, School of Physics and Astronomy, Shanghai Jiao Tong University, Shanghai {\rm 200240}, China}
\address[{\rm16}]{Laboratory for Laser Plasmas and Collaborative Innovation Centre of IFSA, Shanghai Jiao Tong University, Shanghai {\rm 200240}, China}
\address[{\rm17}]{Tencent Youtu Lab, Shanghai {\rm 200030}, China}
\address[{\rm18}]{Guizhou Provincial Key Laboratory of Radio Astronomy and Data Processing, Guizhou Normal University, Guiyang {\rm 550001}, China}
\address[{\rm19}]{Purple Mountain Observatory, Chinese Academy of Sciences, Nanjing {\rm 210023}, China}

\abstract{Fast radio bursts (FRBs) are millisecond-duration cosmic transients whose origin remains elusive. Competing models invoke either earthquake-like processes or flare-like mechanisms. To discriminate between these scenarios, we develop a novel diagnostic, the Pincus-Lyapunov diagram (PLD), to characterize the energetic transients in the stochasticity-chaos phase space. We compile burst sequences from five representative FRBs (FRB 20121102A, FRB 20190520B, FRB 20201124A, FRB 20220912A, and
FRB 20240114A), together with those from magnetar flares (SGR J1550$-$5418, SGR J0501+4516, SGR 1806$-$20, SGR 1900+14, and SGR J1935+2154), pulsar glitches, solar flares, and earthquakes, and map them onto the PLD for comparative analysis. The resulting diagram shows that FRBs occupy a distinct region of the phase space. Specifically, a permutation test reveals a statistically significant difference in the distributions of magnetar flares and pulsar glitches compared to those of repeating FRBs ($p$-value $\simeq 0.05$). To examine whether temporal variations in source activity can shift a repeater’s position in this phase space, we analyze the time evolution of the most prolific repeater, FRB~20240114A. For this repeating FRB, both Pincus Index and Lyapunov Exponent demonstrate statistically stable behaviour over the eight-month observation session, with Augmented Dickey--Fuller tests yielding $p \simeq 1.78\times10^{-3}$ and $9.91\times10^{-3}$, respectively. By assembling the most comprehensive dataset to date, our work indicates that the trigger mechanisms of repeating FRBs are likely to be distinct from those driving magnetar flares, pulsar glitches, solar flares, and earthquakes.}


\PACS{\\}

\maketitle


\begin{multicols}{2}

\section{Introduction}\label{section1}
Fast radio bursts (FRBs) are extremely luminous millisecond-duration radio transients originating at cosmological distances \cite{lorimer2007}. Despite extensive follow-up and theoretical work, the physical mechanism that triggers FRBs is still a subject of active debate in astrophysics \cite{2019ARA&A..57..417C, 2022A&ARv..30....2P, 2023RvMP...95c5005Z, 2026enap....3..372K}. The discovery of FRB 20200428 in association with a Galactic magnetar (SGR J1935+2154) has provided direct evidence that at least some FRBs can be produced by magnetars \cite{CHIME2020a, Bochenek2020, 2020Natur.587...63L}. This landmark event solidified the magnetar paradigm for FRBs, but also raised new questions about how magnetar activity translates into the observed radio bursts.

In magnetar-based FRB models, an underlying trigger mechanism is required to suddenly release the enormous burst energy on millisecond timescales \cite{2023RvMP...95c5005Z}. Some proposed triggers may be of a quake-like nature, often referred to as ``starquakes''. In this scenario, FRBs are produced by crustal cracking on the surface of neutron stars \cite{2018ApJ...852..140W, 2020ApJ...902L..32D,  2021Innov...200152G, 2021ApJ...919...89Y}. Starquake-driven models naturally link FRBs to phenomena such as pulsar glitches, which are sudden jumps in spin rate believed to result from crustal quakes \cite{2022RPPh...85l6901A}, and to the statistical behaviors of earthquakes \cite{2018ApJ...852..140W}. Alternative triggers are more flare-like, wherein the FRB is directly powered by a magnetospheric eruption or magnetic reconnection event in the neutron star's magnetosphere, with or without a crust crack \cite{2010vaoa.conf..129P}. External triggers, such as external events or binary interactions, have also been suggested \cite{2015ApJ...809...24G,2020ApJ...897L..40D, 2022NatCo..13.4382W}. Distinguishing between these models is challenging but crucial.

One promising approach to discriminate among burst mechanisms is to study the statistical properties of repeating FRB activity, analogous to the temporal clustering of earthquakes or flares. If repeaters are driven by neutron-star quakes, their bursts are expected to behave like seismic events and show clustering in both time and energy \cite{2004PhRvL..92j8501C,2008PhRvL.100c8501L}. Alternatively, if FRBs are produced by flaring events, they may occur more randomly in time and lack clear aftershock patterns, although temporally close flares may show a tendency for the second event to have higher energy \cite{2010A&A...511L...2L}.

Existing statistical studies on whether the bursts of FRBs are correlated have reached different conclusions. The results obtained from the statistical analysis of FRB waiting times can be broadly categorised into three distinct groups. Some studies employ correlation-function analysis \cite{2023MNRAS.526.2795T, 2024MNRAS.530.1885T} or fits the cumulative waiting-time distribution with a Weibull function, obtaining a shape parameter smaller than 1 \cite{2018MNRAS.475.5109O, 2020A&A...635A..61O, 2021ApJ...920L..23Z}. A plausible interpretation of this phenomenon is that it provides evidence for the hypothesis that FRBs cluster in time, in a manner analogous to earthquakes. Nevertheless, these studies often overlook the fundamental difference between the waiting-time distributions of FRBs and earthquakes. In particular, FRB waiting times frequently show a pronounced bimodal structure \cite{2021Natur.598..267L, 2022Natur.609..685X, 2022RAA....22l4002Z,2023MNRAS.524..569W, 2023MNRAS.520.2281N,  2023ApJ...955..142Z}, whereas earthquake waiting times are well described by a smooth, single-peaked Weibull distribution \cite{2009PhyA..388..491H, 2015JSeis..19..721C, 2018NatHa..90..823P, 2024SciBu..69.1020Z}. Some other studies instead fit the waiting-time distribution with a lognormal function \cite{2021ApJ...922..115A, 2021Natur.598..267L, 2022Natur.609..685X, 2022MNRAS.515.3577H}, as in the case of solar flares \cite{2019ApJ...884...50V}, suggesting a more random emission behavior. Yet other studies argue that FRB emission is consistent with a Poisson process, thereby implying completely random emission with no intrinsic temporal correlation \cite{2021MNRAS.500..448C, 2023MNRAS.519..666J,  2023ApJ...956...23S, 2025ApJS..276...20Z, 2025ApJ...992..185W}. 

In addition to waiting-time statistics, the energy distribution of FRBs provides another crucial probe of their underlying emission physics. The burst energy of repeaters usually follows a power-law distribution, which may be resembling self-organized criticality (SOC) systems such as magnetar flares, solar radio bursts and earthquakes \cite{2017JCAP...03..023W, 2021MNRAS.501.3155W}. Some high-cadence observations, especially of hyperactive repeaters such as FRB 20121102A, further supported the power-law form but revealed significant deviations at high energies, including apparent cutoffs or breaks \cite{2016MNRAS.461L.122L, 2018MNRAS.481.2320L,2022Natur.609..685X}. Observations show that the burst-energy distribution of several active repeaters, such as FRB 20121102A and FRB 20201124A, displays a pronounced bimodal structure, with the low-energy component closely following a lognormal form and the high-energy component requiring an additional distinct population \cite{2021Natur.598..267L, 2022RAA....22l4002Z}. Meanwhile, population studies of both repeating and apparently non-repeating FRBs indicate that their isotropic-equivalent energies may follow different distributions \cite{2022MNRAS.511.1961H}, raising the possibility that the two classes involve distinct progenitor mechanisms or environmental conditions.

The discordant perspectives derived from these analyses underscore the persistent uncertainty surrounding the true nature of FRB triggering.
They motivate examining FRB variability with multiple diagnostics and comparing it directly with known physical processes, including both seismic and flaring systems, in order to clarify the underlying mechanism. More recently, several works have jointly analyzed burst times and energies to search for possible correlations and clustering patterns in FRB activity. In a specific study, the waiting time between successive bursts and the change in energy from burst to burst show no significant correlation and remain essentially unpredictable, indicating an absence of the kind of time–energy clustering seen in seismic aftershock sequences \cite{2024SciBu..69.1020Z}. Other analyzes further suggest that the statistics of repeating FRBs are difficult to reconcile with those of typical magnetar bursts, implying that the underlying emission process may differ from that of standard magnetar flares \cite{2023SciA....9F6198Z,2024MNRAS.528L.133Y}. These results further motivate a comparative study of FRB statistics.

Comprehensive periodicity searches have been carried out for all active repeaters. The lack of spin-related periodicity poses a challenge to the widely assumed compact object origin of FRBs \cite{2020MNRAS.496.4565C, 2022RAA....22l4004N, 2022Natur.606..873N, 2023MNRAS.520.2281N, 2023MNRAS.519..666J, Zhou2026}. This motivates us to characterize the FRBs beyond just time domain analysis, particularly, periodicity analysis. Adopted from nonlinear dynamics, the Pincus Index ($PI$) measures the degree of randomness in a time series \cite{1991PNAS...88.2297P, 2019NatSR...912761D}. $PI$ measures the ratio between informational entropy before and after a randomization operation. Thus a $PI$ value close to unity means that the original time series is already random. In this work, we generalize the $PI$ to a two-dimensional energy-time parameter space \cite{2024SciBu..69.1020Z}.
We also adopt the classical description of chaos, namely the Lyapunov Exponent ($LE$), which quantifies the sensitivity to initial conditions and thus characterizes its degree of chaotic behavior 
 \cite{2010cfsm.book.....C}.
By computing $PI$ and $LE$ for each FRB sample and for each comparison data set, we map all sources into a two-dimensional ``stochasticity–chaos'' phase space, also known as the Pincus-Lyapunov diagram (PLD). This framework enables us to  quantify the FRB behaviors in terms of time and energy in relation to known phenomena \cite{2024SciBu..69.1020Z, 2024MNRAS.528L.133Y, 2024MNRAS.533..872S}, including earthquakes, solar bursts, magnetar flares, glitches.

In this work, we carry out a PLD analysis of the largest available data sets of repeating FRBs. We compile five extensive samples of repeaters, each FRB source contains a large number of bursts detected over long observing campaigns. We then compare their time–energy statistics with those of magnetar X-ray flares, pulsar glitches, solar flares, and earthquakes. These four classes provide representative examples of flare-like and quake-like phenomena. Magnetar flares act as the magnetospheric analogue of FRB activity, pulsar glitches illustrate stellar starquakes, solar flares trace magnetic reconnection in stellar coronae, and earthquakes serve as the canonical example of self-organized crustal failure. 

In the following sections, we describe the data sets and analysis methods in Section~\ref{sec:data}, present the main results of our study in Section~\ref{sec:Stochasticity and chaos}, discuss the possible temporal evolution of repeating FRBs and the implications for their underlying physical mechanisms in Section~\ref{sec:discussion}, and summarize our conclusions in Section~\ref{sec:conclusion}.

\section{Data and method}\label{sec:data}
To conduct a comprehensive comparative study of the statistical properties of FRBs and other transients, we assemble event sequences that are as statistically rich as possible. All data sets used in this work are summarized in Table~\ref{tab:data}. The data underlying this work are publicly available in the Science Data Bank (ScienceDB; DOI: \href{https://doi.org/10.57760/sciencedb.Fastro.00031}{10.57760/sciencedb.Fastro.00031}).

\begin{table*}
    \centering
    \caption{Summary of data sets \label{tab:data}}
    \scriptsize
    \begin{tabular}{ccccc}
        \hline
        & Source & Observation period & Event number & Reference \\
        \hline
        \multirow{5}{*}{FRB}
        & FRB 20121102A         & Aug 2019-Oct 2019 & 1652   & \cite{2021Natur.598..267L} \\
        & FRB 20190520B         & May 2019-Jul 2021 & 328    & \cite{2023Sci...380..599A} \\
        & FRB 20201124A         & Sep 2021           & 881    & \cite{2022RAA....22l4002Z} \\
        & FRB 20220912A         & Oct 2022-Dec 2022 & 1076   & \cite{2023ApJ...955..142Z} \\
        & FRB 20240114A         & Jan 2024-Aug 2024 & 11073  & \cite{2025arXiv250714707Z} \\
        \hline
        \multirow{7}{*}{Magnetar flare}
        & SGR J1550-5418        & Oct 2008-Apr 2009 & 352    & \cite{2015ApJS..218...11C} \\
        & SGR J0501+4516        & Aug 2008-Sep 2009 & 27     & \cite{2015ApJS..218...11C} \\
        & SGR 1806-20           & Nov 1996-Jun 2011 & 919    & Magnetar Database$^{a}$ \\
        & SGR 1900+14           & Jun 1998-Apr 2006 & 431    & Magnetar Database$^{a}$ \\
        & SGR J1935+2154        & Apr 2020           & 204    & \cite{2020ApJ...904L..21Y} \\
        & SGR J1935+2154        & Oct 2022-Nov 2022 & 268    & \cite{2024ApJ...976...99S} \\
        & SGR J1935+2154 (radio)& Oct 2020           & 563    & \cite{2024ApJS..275...39W} \\
        \hline
        \multirow{4}{*}{Pulsar glitch}
        & PSR B1737-30          & Jul 1987-Dec 2022 & 37     & \cite{2022MNRAS.510.4049B} \\
        & PSR J0537-6910        & Apr 1999-Oct 2023 & 65     & \cite{2022MNRAS.510.4049B} \\
        & PSR B1338-62          & Apr 1990-May 2017 & 33     & \cite{2022MNRAS.510.4049B} \\
        & PSR B0833-45          & Feb 1969-Sep 2021 & 25     & \cite{2022MNRAS.510.4049B} \\
        \hline
        Earthquake & Earthquake & Jan 1932-Apr 2021 & 820132 & SCEDC$^{b}$ \\
        \hline
        Solar flare & Solar flare & Oct 2006-Dec 2020 & 17357 & Hinode Flare catalogue$^{c}$ \\
        \hline
    \end{tabular}
    
    \vspace{4pt} 
    \raggedright
    a) Sabancı University, Magnetar Database, \url{https://magnetars.sabanciuniv.edu}; b) Southern California Earthquake Data Center (SCEDC), \url{scedc.caltech.edu}, doi: \href{https://doi.org/10.7909/C3WD3xH1}{10.7909/C3WD3xH1}. c) Masuda, S., K. Watanabe, and T. Segawa, Hinode Flare Catalogue, doi: \href{https://www.isee.nagoya-u.ac.jp/doi/10.34515/CATALOG_HINODE-00000.html}{10.34515/CATALOG.HINODE-00000, 2021}
\end{table*}

For the FRB sample-set we focus on five repeaters that have been followed with the Five-hundred-meter Aperture Spherical radio Telescope (FAST), namely FRB~20121102A, FRB~20190520B, FRB~20201124A, FRB~20220912A, and FRB~20240114A. All five sources have produced at least a few hundred detected bursts in FAST campaigns, and for FRB~20240114A the burst count exceeds $10^4$, which makes it the most active repeater currently known. For FRB~20121102A and FRB~20190520B, we adopt the published burst event lists of \cite{2024SciBu..69.1020Z, 2024MNRAS.528L.133Y}. The remaining three FRBs, FRB~20201124A \cite{2022RAA....22l4002Z}, FRB~20220912A \cite{2023ApJ...955..142Z}, and FRB~20240114A \cite{2025arXiv250714707Z}, are drawn from recent FAST monitoring campaigns and are selected purely on the basis of data quality and burst statistics, since they provide long, densely sampled sequences with relatively uniform sensitivity and well-calibrated burst energies.

To represent quake-like activity in neutron stars we turn to pulsar glitches. We focus on four sources with prolific and well-characterized glitch histories, namely PSR~B1737$-$30, PSR~J0537$-$6910, PSR~B1338$-$62, and PSR~B0833$-$45 (Vela). 
The parameters of glitches are taken from the Australia Telescope National Facility (ATNF) pulsar catalog \cite{Manchester2005} and the Jodrell Bank glitch catalog \cite{Espinoza2011,2022MNRAS.510.4049B}. These pulsars have experienced dozens of glitches over observational baselines of decades, and for each event the time of occurrence and the jump in spin frequency or spin-down rate are measured with high precision. Consequently, their glitch records form natural time series of crust-failure events that can be compared directly with the FRB trains.

For flare-like phenomena we collect seven burst sequences from five magnetars. Short X-ray bursts of SGR~J1550$-$5418 and SGR~J0501+4516 are taken from the Fermi Gamma-ray Burst Monitor (Fermi-GBM) sample of \cite{2015ApJS..218...11C}. The X-ray bursts of SGR~1806$-$20 and SGR~1900+14 are drawn from Rossi X-ray Timing Explorer (RXTE) observations compiled in the Magnetar Catalog hosted by Sabancı University. For SGR~J1935+2154 we use two NICER burst samples, one from the 2020 active episode and one from the 2022 episode \cite{2020ApJ...904L..21Y,2024ApJ...976...99S}. In addition, we include the radio bursts from SGR~J1935+2154 observed by FAST in 2020 \cite{2023SciA....9F6198Z,2024ApJS..275...39W}, which provide a direct bridge between magnetar radio emission and the repeating FRB sample. Together, these data sets span a wide range of magnetar bursting behavior and serve as our main reference for flare-like processes.

In order to establish a more extensive context for the neutron-star phenomena, we also include two high-statistics samples, namely earthquakes and solar flares. For both samples we follow \cite{2024SciBu..69.1020Z} in our catalog selections and preprocessing. Earthquake data are drawn from the Southern California Earthquake Data Center (SCEDC) catalog, which covers the period from 1932 to 2021 and comprises more than $8\times10^5$ events in a well-studied tectonic region. Solar-flare data are taken from the Hinode Flare Catalogue, which lists more than $1.7\times10^4$ soft X-ray events observed between 2006 and 2020. These two catalogs provide high-statistics benchmarks for self-organized crustal failures on Earth and for magnetically driven flares in the solar corona, respectively.

For each source or catalog in Table~\ref{tab:data}, we construct an ordered sequence $(t_i, E_i)$, where $t_i$ is the occurrence time and $E_i$ is the energy-like amplitude of the $i$th event. For FRBs or the radio bursts of magnetars, we use the isotropic-equivalent burst energies reported in the original publications, in units of erg. For magnetar X-ray bursts, $E_i$ is taken to be a fluence-like quantity: the fluence derived from a Comptonized model (COMP) for SGR~J1550-5418 and SGR~J0501+4516 \cite{2015ApJS..218...11C}; the reported photon counts for SGR~1806-20 and SGR~1900+14 \href{https://magnetars.sabanciuniv.edu}{(Magnetar Database)}; and the flux integrated over the burst duration based on absorbed blackbody or power-law spectral fits for SGR~J1935+2154 \cite{2020ApJ...904L..21Y,2024ApJ...976...99S}.
For pulsar glitches we adopt the measured change in rotational energy as the event amplitude. For earthquakes and solar flares we use energies obtained by converting the cataloged magnitudes and soft X-ray classes using the empirical relations. Specifically, for earthquakes we use the empirical formula $\log_{10} E = 12.24 + 1.44 M$ from \cite{Bath1966} to convert magnitudes to energies. For solar flares we use the relationship given by the NOAA/GOES XRS classification\footnote{\url{https://www.swpc.noaa.gov/phenomena/solar-flares-radio-blackouts}} to convert flare levels to energies.

These observables increase monotonically with the physical energy released. This property is sufficient for our purposes, because we are interested in the relative statistics of event sizes and their correlations with occurrence times rather than in the absolute energetics. The resulting preparation yields, for every FRB, magnetar, pulsar glitch, earthquake, and solar-flare catalog, a one-dimensional sequence of event times ${t_i}$ and corresponding amplitudes ${E_i}$. We then construct the differences between successive events, defined as the waiting time $\Delta t_i = t_{i+1}-t_i$ and the energy change $\Delta E_i = E_{i+1}-E_i$. The burst sequence is divided into different sessions according to the observation. To remove scale dependence, both sequences are normalized to the interval $[0,1]$.

To quantify the dynamical properties of these sequences, we employ two diagnostics from nonlinear time-series analysis, namely the $PI$ and the $LE$. Physically, they characterize two distinct forms of unpredictability. $PI$ quantifies the absence of repeating patterns in the sequence, and a high $PI$ corresponds to a memoryless, noise-like process whose unpredictability is statistically stable in time. $LE$ quantifies the divergence rate of nearby trajectories in phase space, and a positive $LE$ indicates deterministic sensitivity to initial conditions whose unpredictability grows exponentially in time. The joint $(PI,LE)$ position therefore acts as a diagnostic of the underlying trigger mechanism.

The $PI$ measures the degree of randomness in a sequence, which is based on approximate entropy (ApEn). ApEn measures the regularity of a time series by evaluating the similarity of patterns in the data \cite{1991PNAS...88.2297P}. For a sequence $X=\{x_1, x_2, \dots, x_N\}$, embedding vectors of dimension $m$ are constructed and compared within a tolerance $r$. Here, embedding vectors are constructed to reconstruct the phase space of a dynamical system from a single scalar time series. For an embedding dimension $m$, the $i$-th embedding vector is formed from $m$ consecutive samples starting at index $i$, namely $[x_i, x_{i+1}, \dots, x_{i+m-1}]$. This reconstruction enables similarity comparison between system states and serves as the foundation for both the Approximate Entropy and the Lyapunov Exponent calculations. The ApEn is estimated as:

\begin{equation}
\mathrm{ApEn}(m,r,N)
=
-\frac{1}{N-m}
\sum_{i=1}^{N-m}
\ln
\left(
\frac{C_i^{m+1}(r)}{C_i^{m}(r)}
\right),
\end{equation}
where $C_i^{m}(r)$ denotes the fraction of vectors that remain within a distance $r$ of the $i$th vector in an $m$-dimensional embedding space.

To reduce the dependence on the tolerance parameter, the Max Approximate Entropy (MaxApEn) is introduced:

\begin{equation}
\mathrm{MaxApEn}
=
\max_{r}
\left|
\mathrm{ApEn}(m+1,r) - \mathrm{ApEn}(m,r)
\right|.
\end{equation}

Following \cite{2019NatSR...912761D}, the Pincus Index is defined as the ratio between the MaxApEn of the original sequence and that of shuffled sequence generated by Monte Carlo simulation.
\begin{equation}
{PI}
=
\frac{\mathrm{MaxApEn}_{\mathrm{original}}}{\mathrm{MaxApEn}_{\mathrm{shuffled}}}.
\end{equation}

Thus, smaller $PI$ values indicate stronger regularity and predictability, while larger values correspond to more stochastic behavior. Specifically, if the $PI$ approaches $1$, it implies that the original sequence is already highly disordered and almost completely random. Consequently, its entropy becomes nearly indistinguishable from that of the randomly shuffled data.

We compute the $PI$ separately from the normalized waiting-time sequence and energy-difference sequence, yielding ${PI}_t$ and ${PI}_E$. The final $PI$ of the source is defined as:
\begin{equation}
{PI} = \frac{{PI}_t + {PI}_E}{2}.
\end{equation}

The $LE$ is a measure of the dynamical stability of a system, which quantifies the sensitivity of its trajectories to initial conditions over long-term evolution. For two trajectories in phase space with an initial separation vector $\delta_0$, the divergence evolves approximately as:

\begin{equation}
|\delta(t)| \simeq |\delta_0| e^{\lambda t},
\end{equation}
where $\lambda$ is the Lyapunov Exponent. 

The maximum Lyapunov Exponent (MLE) is the largest exponent of the system and characterizes the overall stability of the dynamics. Negative MLE values indicate stable or periodic behavior, while positive values imply chaotic dynamics \cite{2010cfsm.book.....C}. In this work, the MLE is estimated using the \texttt{NOLDS} package \cite{scholzel2019nonlinear}. Specifically, we use the \texttt{lyap\_e} function, which implements the algorithm of Eckmann et al.\ \cite{Eckmann1986}. The algorithm reconstructs the phase space from the embedding vectors, identifies local neighborhoods for each state, estimates the Jacobian matrices governing their evolution, and computes the full Lyapunov spectrum via QR decomposition.

The $LE$ is computed separately from the normalized waiting-time and energy sequences, yielding $\lambda_t$ and $\lambda_E$. The final Lyapunov index is defined as:

\begin{equation}
{LE} = \frac{\lambda_t + \lambda_E}{2}.
\end{equation}

For each source, the waiting-time and energy sequences are embedded in phase space and the $(PI,LE)$ pair is computed using standard algorithms. Each source is therefore represented by a point in the stochasticity–chaos phase space, which enables a direct comparison between repeating FRBs and other transients, including magnetar flares, pulsar glitches, earthquakes, and solar flares.

\section{Results}\label{sec:Stochasticity and chaos}

\subsection{The Pincus-Lyapunov diagram}\label{subsec:PLD}

\begin{figure*}
    \centering
    \includegraphics[width=0.8\textwidth]{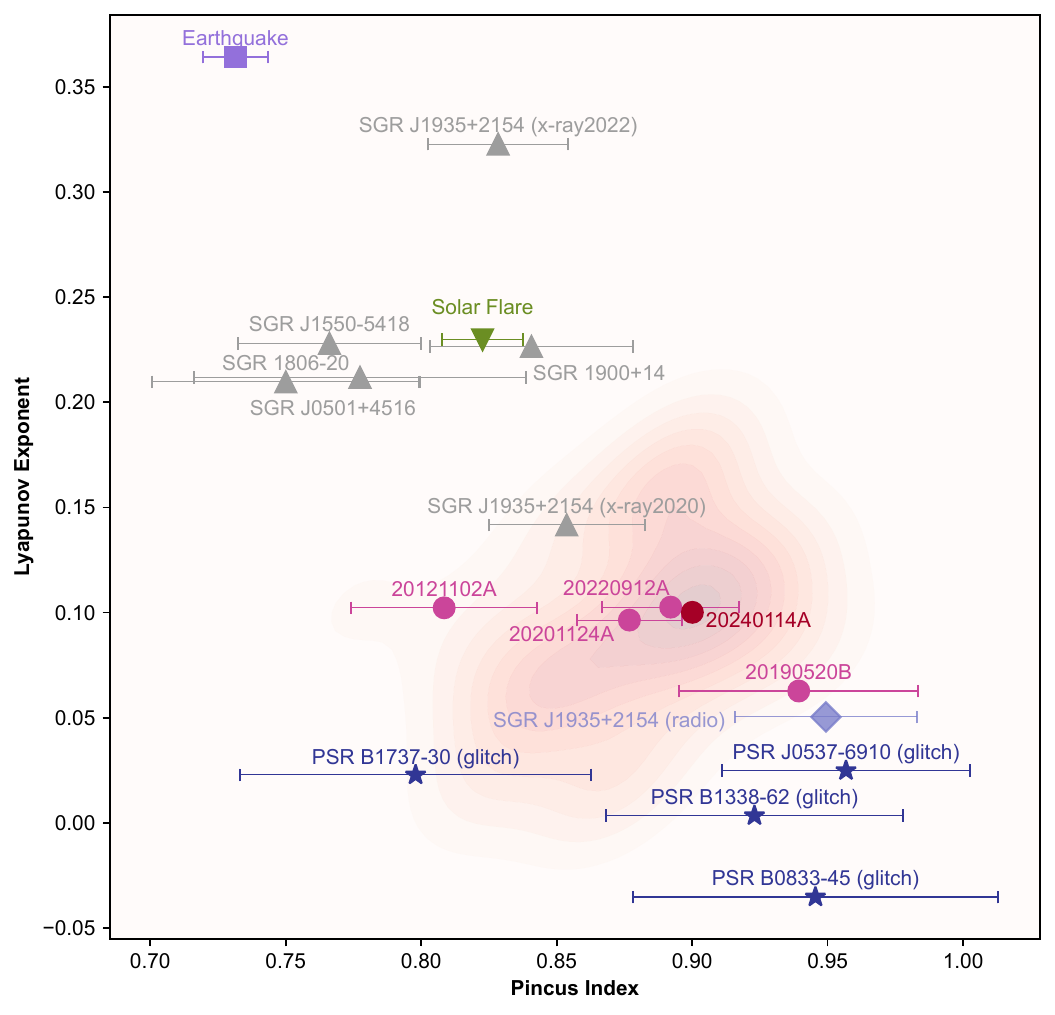}
     \caption{Pincus–Lyapunov diagram of transients. The purple squares, grey triangles, green inverted triangles, red circles (including light red and dark red circles), soft violet diamonds, and blue stars represent earthquakes, magnetar X-ray flares, solar flares, repeating FRBs,magnetar radio pulses,  and  events of pulsar glitches, respectively. Horizontal error bars indicate the standard deviation of $PI$ for each source. The red shaded region shows the kernel density estimate of individual observing sessions of FRB~20240114A in the $PI$–$LE$ plane, with the red point marking the density maximum.}
    \label{fig:PI-LE}
\end{figure*}

Mapping all event sequences onto the Pincus–Lyapunov diagram (PLD) reveals clear structure in the stochasticity–chaos phase space. Figure~\ref{fig:PI-LE} shows $PI$ versus $LE$ for earthquakes, magnetar X-ray flares, solar flares, repeating FRBs, and pulsar glitches. Rather than scattering randomly, the sources align along a broad diagonal band from high chaos and low stochasticity (top left) to low chaos and high stochasticity (bottom right).

Earthquakes occupy the top-left part of the PLD, indicating strong chaotic behavior and relatively low stochasticity. At the opposite end, glitches cluster at very high $PI$ and low $LE$, corresponding to highly stochastic but only weakly chaotic sequences. Magnetar flares, solar flares, and FRBs fall between these two extremes. Overall, the degree of chaos decreases roughly in the order earthquakes $>$ magnetar flares $\simeq$ solar flares $>$ FRBs $>$ pulsar glitches, while the level of stochasticity follows the opposite trend.

Phenomena of the same class tend to cluster in the diagram, suggesting that the PLD captures intrinsic properties of the underlying physical process. Solar flares largely overlap with magnetar flares, consistent with the view that both are powered by magnetic reconnection in highly magnetized plasmas. In contrast, the FRB points form a compact group that is clearly offset from the main loci of magnetar flares and glitches. Repeating FRBs thus behave neither like a straightforward extension of solar or magnetar flares to radio wavelengths, nor like a direct analogue of starquake-driven glitches, at least in terms of their time–energy statistics.

The Galactic magnetar SGR~J1935+2154 provides a particularly instructive case. Its April 2020 active episode produced the first detected Galactic FRB, FRB~20200428, in association with a bright X-ray flare. In Figure~\ref{fig:PI-LE}, the points labeled ``SGR J1935+2154 (X-ray 2020)'' and ``SGR J1935+2154 (radio)'' represent the X-ray and radio bursts after this episode, respectively. While ``SGR J1935+2154 (X-ray 2022)'' corresponds to the later 2022 X-ray activity without a reported FRB counterpart. Both the 2020 X-ray bursts and the associated radio bursts lie very close to the FRB cluster in the PLD, providing independent statistical support for the physical link between this magnetar outburst and the FRB mechanism. By contrast, the 2022 X-ray bursts fall well away from the FRB region and align more closely with the general magnetar-flare population. The change in location between the 2020 and 2022 episodes suggests that a single source can operate in distinct dynamical modes. This contrast also disfavors a band-sensitivity interpretation of the PLD. The 2020 and 2022 points share the same X-ray band, yet sit in different regions of the diagram, suggesting that the offset reflects intrinsic source dynamics rather than the wavelength of detection. We nevertheless do not exclude the potential influence of observational uncertainties and individual outliers, given the limited sample sizes and the diversity of detection conditions across these episodes.

\subsection{Statistical distinction}

To quantify the similarity between repeating FRBs and other populations, a geometric distance in the two-dimensional PLD is introduced. The kernel density estimate (KDE) peak of the five FRB sources, $(PI_0, LE_0) \simeq (0.90, 0.09)$, is adopted as the reference point. For source $i$, the Euclidean distance is defined as:

\begin{equation}
    d_{\mathrm{i}}=\sqrt{\left(P I_{\mathrm{i}}-P I_0\right)^2+\left(L E_{\mathrm{i}}-L E_0\right)^2}.
\end{equation}

A smaller $d_i$ indicates that the stochastic and chaotic properties of the source are closer to those of a typical FRB repeater. For FRB~20240114A, the most active source in our sample, we further divide the burst sequence into 40 segments and compute $(PI, LE)$ for each segment. This allows us to treat FRB~20240114A as an ensemble of quasi-independent realizations, and to compare its internal variability with the scatter among different FRB sources.

\begin{figure}[H]
    \centering        
    \includegraphics[scale=0.6]{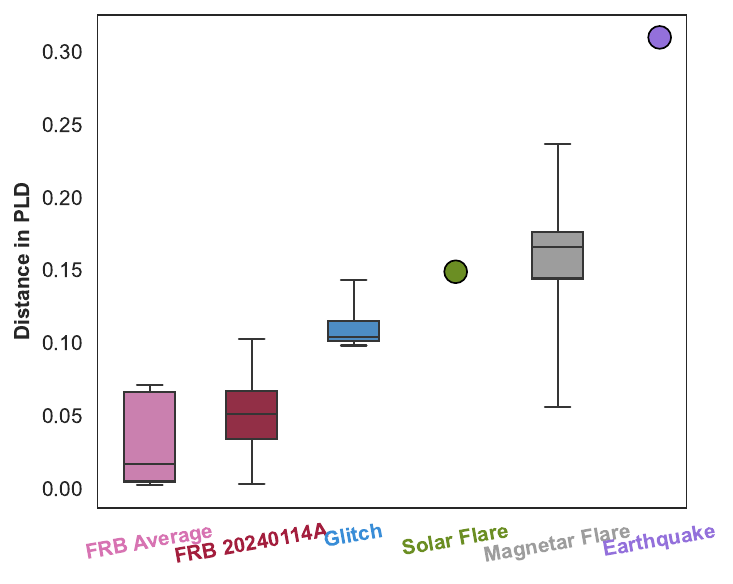}
    \caption{Box plots of the Euclidean distance to the FRB centroid in the PLD.  For each class, the box spans the interquartile range, the horizontal line inside the box marks the median, and the whiskers extend to the extrema of the distribution.}
    \label{fig:boxplot}
\end{figure}

The distribution of distances is shown in Figure~\ref{fig:boxplot}. The average FRB sample, and the 40 segments of FRB~20240114A considered separately, both lie very close to the FRB centroid, confirming the internal consistency of the FRB class and indicating that FRB~20240114A is statistically representative despite its extreme activity level. Glitches, solar flares, magnetar flares and earthquakes have larger distances. This visual impression suggests that typical magnetar flares and earthquakes occupy a region of the PLD that is systematically different from that of repeating FRBs, while glitches and solar flares form a nearer but still distinct population.

To convert this qualitative separation into a formal statistical statement, we test whether the FRB points in the PLD could plausibly be drawn from the same parent distribution as the other populations. For this purpose, we use the ``energy distance'' as a non-parametric measure of discrepancy between two multivariate distributions, combined with a permutation test to evaluate statistical significance. Conceptually, the energy distance compares the pairwise Euclidean distances between data points from different samples with those computed within each individual sample. If the samples $X$ and $Y$ are well mixed in the PLD, the cross-sample distances are expected to be comparable to the within-sample distances, resulting in a small energy distance. If the samples come from distinct clusters, cross-sample distances will on average be larger, leading to a greater energy distance.

Formally, let $\mathbf{X}$ and $\mathbf{Y}$ be random vectors in the PLD with cumulative distribution functions $F$ and $G$. The energy distance between $F$ and $G$ is
\begin{equation}\label{eq:energy-distance}
    T(F,G) = 2 \mathbb{E}\bigl\|\mathbf{X} - \mathbf{Y} \bigr\|
             - \mathbb{E}\bigl\|\mathbf{X} - \mathbf{X}'\bigr\|
             - \mathbb{E}\bigl\|\mathbf{Y} - \mathbf{Y}'\bigr\|.
\end{equation}
where $\mathbf{X}'$ and $\mathbf{Y}'$ are independent copies of $\mathbf{X}$ and $\mathbf{Y}$, and $\|\cdot\|$ denotes the Euclidean norm. The statistic $T(F,G)$ is non-negative and equals zero only when $F=G$, larger values therefore indicate stronger evidence that the two distributions differ.

To evaluate whether an observed value $T_{\rm obs}$ is unusually large under the null hypothesis $H_0: F=G$, we perform a permutation test. We pool the FRB sample with the comparison sample, randomly shuffle the labels, and recompute the energy distance for each random relabeling. This procedure generates a reference distribution $\{T_{{\rm perm},i}\}$ that approximates the sampling distribution of $T$ under $H_0$ without assuming any particular functional form for $F$ or $G$. The $p$-value is then given by the fraction of permutations in which the randomized statistic exceeds the observed one.

\begin{equation}\label{eq:pvalue}
    p = \frac{1}{N_{\mathrm{perm}}}
        \sum_{i=1}^{N_{\mathrm{perm}}} \mathbf{1}\left(T_{\mathrm{perm},i} \ge T_{\mathrm{obs}}\right).
\end{equation}
where $N_{\mathrm{perm}}$ is the total number of random permutations, for which we adopt $N_{\mathrm{perm}} = 10{,}000$ in this work, and $\mathbf{1}(\cdot)$ is the indicator function. A small $p$-value implies that the observed separation is unlikely to arise by chance if the two samples were drawn from the same distribution.

The resulting energy distances and $p$-values for comparisons with the FRB population are listed in Table~\ref{tab:p-value}. For magnetar flares we obtain an energy distance of 0.138 and a $p$-value of 0.031; for pulsar glitches the corresponding values are 0.111 and 0.020. Both $p$-values lie below the standard significance threshold $\alpha=0.05$, so the null hypothesis that magnetar flares (or pulsar glitches) and FRBs share the same parent distribution in the PLD can be rejected at the 5\% significance level. In other words, the stochastic–chaotic signatures of repeating FRBs differ significantly from those of magnetar flares and pulsar glitches. By contrast, when we compare FRB~20240114A with the rest of the FRB sample, the energy distance is only 0.012 and the $p$-value is 0.630, indicating no statistically significant difference. A self-consistency check in which the FRB sample is randomly split into two subsamples yields an energy distance consistent with zero and a high $p$-value of 0.984, as expected.

\begin{table}[H]
    \centering
    \caption{Energy distance and p-value between FRBs and other phenomena.\label{tab:p-value}}
    \scriptsize
    \begin{tabular}{lcc}
        \hline
        Comparison Group & Energy Distance & $p$-value \\
        \hline
        Magnetar Flares        & 0.138 & 0.031 \\
        Pulsar Glitches        & 0.111 & 0.020 \\
        FRB 20240114A          & 0.012 & 0.630 \\
        FRB (Self-consistency) & 0.000 & 0.984 \\
        \hline
    \end{tabular}
\end{table}

Taken together, the PLD, the distance distributions, and the permutation tests show that repeating FRBs occupy a distinct region of the stochasticity–chaos phase space. Their time–energy statistics cannot be reproduced by typical magnetar flares or by pulsar glitches, even though all three phenomena are likely associated with highly magnetized neutron stars. At the same time, the internal consistency of the FRB group, including the hyper-active source FRB~20240114A, suggests that repeaters form a coherent class with a characteristic stochastic–chaotic signature.

\section{Discussion}\label{sec:discussion}

\subsection{Temporal stability of FRB~20240114A in PLD}

FRB~20240114A was first identified as a repeating fast radio burst by CHIME in 2024 \cite{2024ATel16420....1S}. Subsequent follow-up observations with Parkes/Murriyang, Westerbork~RT1, the Northern Cross radio telescope, GMRT, and several other facilities confirmed its repeating nature and exceptionally high activity \cite{2024ATel16430....1U,2024ATel16432....1O, 2024ATel16434....1P, 2024ATel16452....1K}. 

Motivated by its extreme activity, FAST initiated systematic monitoring of FRB~20240114A in late January 2024. These observations revealed burst rates that vary dramatically with time, with peak instantaneous rates reaching up to 729~bursts~h$^{-1}$, making FRB~20240114A among the most active repeaters known. By August 2024, FAST alone had detected more than $10^{4}$ bursts from this source \cite{2025arXiv250714707Z}, while the burst rate varied by orders of magnitude over months.

\begin{figure*}
    \centering
    \includegraphics[width=0.9\textwidth]{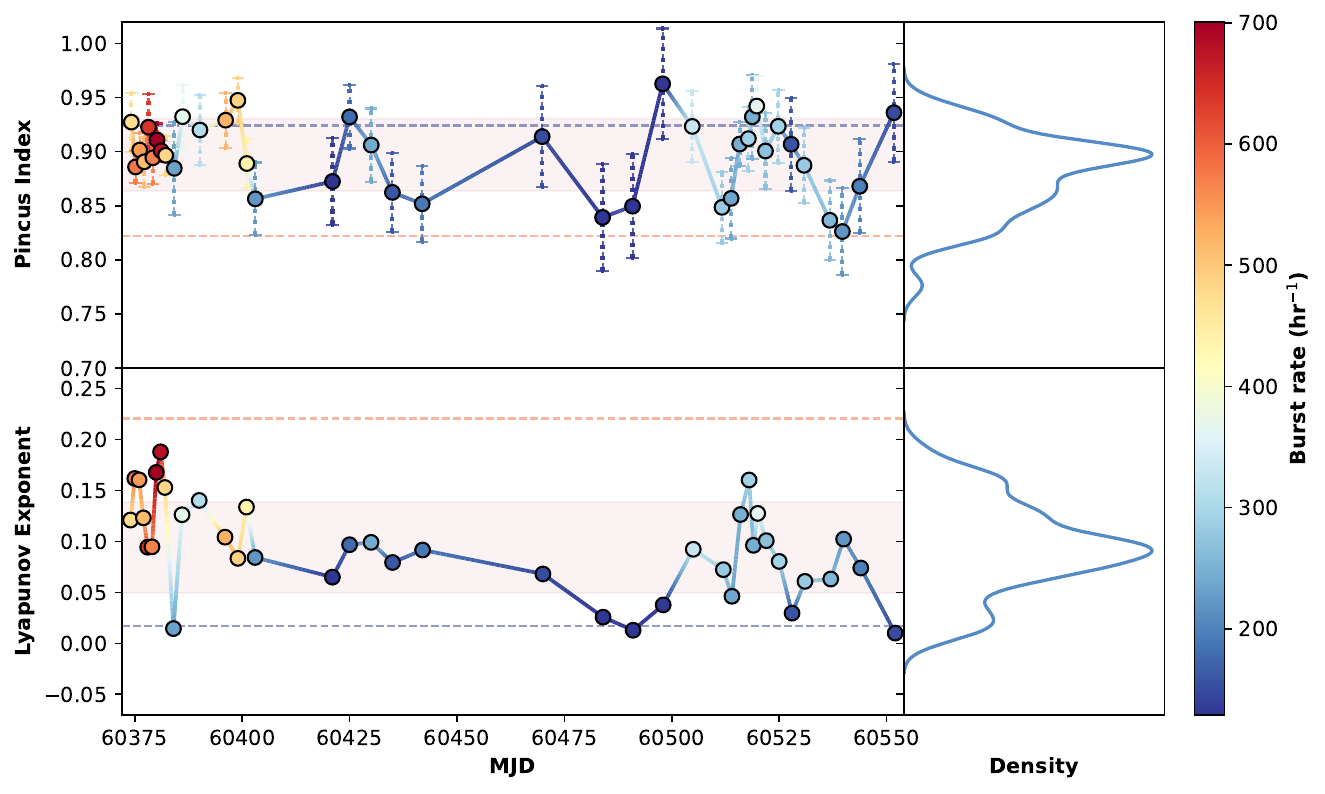}
    \caption{Temporal evolution of $PI$ (upper panel) and $LE$ (lower panel) for FRB~20240114A between MJD~60374 and 60552. Each point represents a single observing session and is colored by the session-averaged burst rate, as indicated by the color bar. The curves on the right show kernel density estimates (KDE) of the marginal distributions of $PI$ and $LE$ over all sessions. Vertical error bars indicate the standard deviation of $PI$. The red shaded regions indicate the $\pm 1\sigma$ ranges relative to the mean values. Orange and blue dashed horizontal lines mark KDE peak values associated with magnetar flares and glitches, respectively.}
    \label{fig:frb20240114A_PILE}
\end{figure*}

To test whether large activity variations can shift a repeater’s position in the PLD, the $PI$ and $LE$ are computed for each individual observing session. The two-dimensional KDE of all sessions of FRB~20240114A is shown in Figure~\ref{fig:PI-LE}, while the temporal evolution is presented in Figure~\ref{fig:frb20240114A_PILE}. The upper panel displays $PI$ as a function of MJD, and the lower panel shows $LE$. Each point corresponds to an observing session, with the color indicating the session-averaged burst rate. 

Both diagnostics exhibit only modest scatter and no apparent secular drift across the full monitoring interval. Over nearly eight months, $PI$ fluctuates within a narrow range around $0.90$, and LE remains close to $0.10$. Importantly, high-rate segments (warm colors) are interspersed with low-rate segments (cool colors) at similar $PI$ and $LE$ values, with no monotonic trend in either panel. Visually, despite the burst rate varying by more than an order of magnitude, neither diagnostic shows a clear long-term evolution or correlation with the instantaneous activity level.

Furthermore, comparison with the reference population of other transient phenomena in Figure~\ref{fig:frb20240114A_PILE} reveals systematic offsets in the PLD. For magnetar flares, the KDE peaks of both $PI$ and $LE$ lie outside the $\pm 1\sigma$ range of FRB~20240114A. For glitches, the KDE peak in $PI$ falls marginally within the $\pm 1\sigma$ interval near its boundary, whereas the corresponding $LE$ peak remains outside this range. Taken together, these results indicate that the distribution of FRB~20240114A in the joint $PI$–$LE$ parameter space is largely distinct from that associated with magnetar flares and only partially overlaps with that of glitches.

The apparent stability of these time series is quantified using the Augmented Dickey–Fuller (ADF) test. For both $PI$ and $LE$, the null hypothesis of a unit root is rejected at the 1\% significance level, with $p \simeq 1.78\times10^{-3}$ for the $PI$ series and $p \simeq 9.91\times10^{-3}$ for the $LE$ series. This indicates that both sequences are statistically stationary over the current observing window.

Despite burst-rate variations of orders of magnitude, the $(PI,LE)$ values of FRB~20240114A show no corresponding increase in scatter. For all observing sessions, the derived pairs cluster near the FRB population centroid in the PLD. This behavior indicates that the underlying emission process remains stable on month-long timescales. Changes in activity appear to modulate only the number of bursts, while the statistical character of the time–energy sequence remains nearly unchanged.

\subsection{Astrophysical implications}

Repeating FRBs occupy a region of the PLD that is clearly separated from both magnetar flares and pulsar glitches, and this has direct implications for their trigger mechanism. In the $(PI,LE)$ plane, FRBs show systematically high $PI$ and low $LE$. In other words, their burst sequences are highly stochastic but only weakly chaotic, in sharp contrast to magnetar X-ray bursts and quake-like glitch activity. If repeating FRBs were predominantly triggered by a single well-defined mechanism such as an isolated magnetic reconnection episode or a discrete crust-cracking event, their PLD localization would be expected to overlap more closely with those of flare-like or quake-like populations. The observed offset therefore disfavors a simple one-to-one identification between repeaters and any of these standard transient families.

The high-$PI$, low-$LE$ signature of FRBs observed in our analysis points to a high degree of randomness in their activity. This intrinsic stochasticity is consistent with the absence of short-period modulations in burst sequences, despite extensive searches for periodic signals associated with stellar spin \cite{2021Natur.598..267L, 2022RAA....22l4004N, 2023MNRAS.519..666J, 2023MNRAS.520.2281N, Zhou2026}.
Furthermore, the energy structure function analysis by \cite{2026MNRAS.545f2043C} implies that FRBs arise from independent stochastic events, distinct from the correlated crustal failures characteristic of earthquakes. Regarding the connection to magnetars, \cite{2024MNRAS.528L.133Y} previously noted that the complexity measures of magnetar bursts do not match those of FRB repeaters. Our work further substantiates this distinction. By leveraging a larger, more diverse sample and unique statistical diagnostics, we confirm that the segregation between magnetar flares and repeating FRBs in the PLD is not an artifact of limited sampling or a specific source peculiarity, but rather reflects a fundamental difference at the population level.

Taken together, the distinct localization of FRBs in the PLD and its temporal stability imply that repeaters are not merely radio analogs of magnetar X-ray flares, pulsar glitches, or solar flares. This comparison alone does not rule out earthquake-like or starquake-driven triggering. Viewing-geometry effects can reshape the observed energy statistics even when the intrinsic dynamics remain quake-like, while spin periodicity may be suppressed below detectability \cite{2025ApJ...988...62L}. The observed dynamics are more naturally explained by a noisy, high-entropy engine. In this scenario, energy is likely dissipated through highly stochastic mechanisms, such as strongly turbulent plasma processes, multiple quasi-independent magnetospheric sites, or random external triggers (e.g., collisions with interstellar objects), with only modest nonlinear coupling between events \cite{2016ApJ...829...27D,2020ApJ...900L..21Y,2021Univ....7...56L,2022ApJ...932L..20M}. The central engine is still likely a compact object such as a magnetar, with its radio emission governed by a stochastic environment, whether intrinsic or externally driven, not by a single coherent oscillator or a purely deterministic global crust-failure cycle \cite{2023RvMP...95c5005Z}.

\section{Conclusion}\label{sec:conclusion}

In this work, we assembled a statistically rich, multi-phenomenon data set of transients and quantified their stochasticity and chaoticity using the $PI$ and $LE$. Our main conclusions are as follows.

First, the Pincus–Lyapunov diagram reveals that different transients occupy distinct regions in the stochasticity–chaos plane. Earthquakes lie in a region of relatively low stochasticity and high chaos; magnetar and solar flares cluster in an intermediate zone consistent with magnetic reconnection; pulsar glitches form a relatively high stochasticity and low chaos  group; and repeating FRBs occupy a unique locus clearly separated from all other classes.

Second, statistical tests confirm that the FRB population is intrinsically distinct. An energy–distance statistic combined with permutation tests shows that the $(PI,LE)$ distribution of FRBs differs significantly from those of magnetar flares and pulsar glitches, while internal FRB subsamples remain statistically consistent with one another. In particular, different temporal segments of the exceptionally active repeater FRB~20240114A all remain within the FRB region of the PLD, despite large variations in burst rate over months. This indicates that the measured statistical properties reflect an intrinsic dynamical pattern rather than short-term rate fluctuations, observational incompleteness, or finite-sample effects.

Finally, the distinct locus of FRBs in the PLD provides insight into their underlying dynamics. The combination of relatively high $PI$ and low $LE$ suggests that repeating FRBs are governed by a process that is strongly stochastic but only weakly chaotic. Burst times and energies exhibit little deterministic memory of previous events and lack the sensitive dependence on initial conditions characteristic of classical chaotic systems. This behavior disfavors simple scenarios in which FRBs are a direct radio analogue of ordinary magnetar X-ray flares or of crustal failure events such as starquakes and glitches. Instead, it is more consistent with emission regulated by a noisy, high-entropy environment around a relatively stable compact source, plausibly a magnetar.

 \Acknowledgements{This work is supported by the National Natural Science Foundation of China under grants 12588202 and 12522305. Yong-Kun Zhang is supported by the Postdoctoral Fellowship Program and China Postdoctoral Science Foundation under Grant Number BX20250158, and the National Natural Science Foundation of China (Grant No. 12573096). Pei Wang acknowledges support from the CAS Youth Interdisciplinary Team, the Youth Innovation Promotion Association CAS (id. 2021055), and the Cultivation Project for FAST Scientific Payoff and Research Achievement of CAMS-CAS. Di Li is a New Cornerstone Investigator. Yong-Feng Huang is supported by the National Key R\&D Program of China (2021YFA0718500) and by the Xinjiang Tianchi Program. Ru-Shuang Zhao is supported by the National Natural Science Foundation of China (Grant No. 12563008), Science and Technology Foundation of Guizhou Provincial Department of Education (No. KY(2023)059), Liupanshui Science and Technology Development Project (No. 52020-2024-PT-01).\\[\baselineskip]
\textbf{Conflict of interest}\quad The authors declare that they have no conflict of interest.}



\def \apj {ApJ}
\def \apjl {ApJL}
\def \aap {A\&A}
\def \atel {The Astronomer's Telegram}
\def \apjs {ApJS}

\end{multicols}


\begin{thebibliography}{99}

\bibitem{lorimer2007} D. R. Lorimer, M. Bailes, and M. A. McLaughlin et al., Science, 318, 777 (2007).

\bibitem{2019ARA&A..57..417C} J. M. Cordes and S. Chatterjee, ARA\&A, 57, 417 (2019).

\bibitem{2022A&ARv..30....2P} E. Petroff, J. W. T. Hessels, and D. R. Lorimer, Astron. Astrophys. Rev. 30, 2 (2022).

\bibitem{2023RvMP...95c5005Z} B. Zhang, Rev. Mod. Phys. 95, 035005 (2023).

\bibitem{2026enap....3..372K} J. I. Katz, in \textit{Encyclopedia of Astrophysics}, Vol. 3, edited by I. Mandel, A. King, and F. van Leeuwen (Elsevier, Amsterdam, 2026), pp. 372-382.

\bibitem{CHIME2020a} CHIME/FRB Collaboration, B. C. Andersen, and K. M. Bandura et al., Nature, 587, 54 (2020).

\bibitem{Bochenek2020} C. D. Bochenek, V. Ravi, and K. V. Belov et al., Nature, 587, 59 (2020).

\bibitem{2020Natur.587...63L} L. Lin, C. F. Zhang, and P. Wang et al., Nature, 587, 63 (2020).

\bibitem{2018ApJ...852..140W} W. Wang, R. Luo, and H. Yue et al., \apj, 852, 140 (2018).

\bibitem{2020ApJ...902L..32D} C. Dehman, D. Viganò, and N. Rea et al., \apjl, 902, L32 (2020).

\bibitem{2021Innov...200152G} J. Geng, B. Li, and Y. Huang, The Innovation 2, 100152 (2021).

\bibitem{2021ApJ...919...89Y} Y.-P. Yang and B. Zhang, \apj, 919, 89 (2021).

\bibitem{2022RPPh...85l6901A} D. Antonopoulou, B. Haskell, and C. M. Espinoza, Rep. Prog. Phys. 85, 126901 (2022).

\bibitem{2010vaoa.conf..129P} S. B. Popov and K. A. Postnov, in \textit{Evolution of Cosmic Objects through their Physical Activity}, edited by H. Harutyunian, A. Mickaelian, and Y. Terzian (Armenian Astronomical Society, Yerevan, 2010), pp. 129-132.

\bibitem{2015ApJ...809...24G} J. J. Geng and Y. F. Huang, \apj, 809, 24 (2015).

\bibitem{2020ApJ...897L..40D} Z. G. Dai, \apjl, 897, L40 (2020).

\bibitem{2022NatCo..13.4382W} F. Y. Wang, G. Q. Zhang, and Z. G. Dai et al., Nat. Commun. 13, 4382 (2022).

\bibitem{2004PhRvL..92j8501C} Á. Corral, Phys. Rev. Lett. 92, 108501 (2004).

\bibitem{2008PhRvL.100c8501L} E. Lippiello, L. de Arcangelis, and C. Godano, Phys. Rev. Lett. 100, 038501 (2008).

\bibitem{2010A&A...511L...2L} E. Lippiello, L. de Arcangelis, and C. Godano, \aap, 511, L2 (2010).

\bibitem{2023MNRAS.526.2795T} T. Totani and Y. Tsuzuki, MNRAS, 526, 2795 (2023).

\bibitem{2024MNRAS.530.1885T} Y. Tsuzuki, T. Totani, and C.-P. Hu et al., MNRAS, 530, 1885 (2024).

\bibitem{2018MNRAS.475.5109O} N. Oppermann, H.-R. Yu, and U.-L. Pen, MNRAS, 475, 5109 (2018).

\bibitem{2020A&A...635A..61O} L. C. Oostrum, Y. Maan, and J. van Leeuwen et al., \aap, 635, A61 (2020).

\bibitem{2021ApJ...920L..23Z} G. Q. Zhang, P. Wang, and Q. Wu et al., \apjl, 920, L23 (2021).

\bibitem{2021Natur.598..267L} D. Li, P. Wang, and W. W. Zhu et al., Nature, 598, 267 (2021).

\bibitem{2022Natur.609..685X} H. Xu, J. R. Niu, and P. Chen et al., Nature, 609, 685 (2022).

\bibitem{2022RAA....22l4002Z} Y.-K. Zhang, P. Wang, and Y. Feng et al., Res. Astron. Astrophys. 22, 124002 (2022).

\bibitem{2023MNRAS.524..569W} Y.-B. Wang, A. Kurban, and X. Zhou et al., MNRAS, 524, 569 (2023).

\bibitem{2023MNRAS.520.2281N} K. Nimmo, J. W. T. Hessels, and M. P. Snelders et al., MNRAS, 520, 2281 (2023).

\bibitem{2023ApJ...955..142Z} Y.-K. Zhang, D. Li, and B. Zhang et al., \apj, 955, 142 (2023).

\bibitem{2009PhyA..388..491H} T. Hasumi, T. Akimoto, and Y. Aizawa, Physica A, 388, 491 (2009).

\bibitem{2015JSeis..19..721C} A. Charpentier and M. Durand, J. Seismol. 19, 721 (2015).

\bibitem{2018NatHa..90..823P} S. Pasari and O. Dikshit, Nat. Hazards 90, 823 (2018).

\bibitem{2024SciBu..69.1020Z} Y.-K. Zhang, D. Li, and Y. Feng et al., Sci. Bull. 69, 1020 (2024).

\bibitem{2021ApJ...922..115A} K. Aggarwal, D. Agarwal, and E. F. Lewis et al., \apj, 922, 115 (2021).

\bibitem{2022MNRAS.515.3577H} D. M. Hewitt, M. P. Snelders, and J. W. T. Hessels et al., MNRAS, 515, 3577 (2022).

\bibitem{2019ApJ...884...50V} C. Verbeeck, E. Kraaikamp, and D. F. Ryan et al., \apj, 884, 50 (2019).

\bibitem{2021MNRAS.500..448C} M. Cruces, L. G. Spitler, and P. Scholz et al., MNRAS, 500, 448 (2021).

\bibitem{2023MNRAS.519..666J} J. N. Jahns, L. G. Spitler, and K. Nimmo et al., MNRAS, 519, 666 (2023).

\bibitem{2023ApJ...956...23S} K. R. Sand, D. Breitman, and D. Michilli et al., \apj, 956, 23 (2023).

\bibitem{2025ApJS..276...20Z} Y.-K. Zhang, D. Li, and Y. Feng et al., \apjs, 276, 20 (2025).

\bibitem{2025ApJ...992..185W} X.-W. Wang, Z. Yan, and Z.-Q. Shen et al., \apj, 992, 185 (2025).

\bibitem{2017JCAP...03..023W} F. Y. Wang and H. Yu, J. Cosmol. Astropart. Phys. 2017, 023 (2017).

\bibitem{2021MNRAS.501.3155W} F. Y. Wang, G. Q. Zhang, and Z. G. Dai, MNRAS, 501, 3155 (2021).

\bibitem{2016MNRAS.461L.122L} W. Lu and P. Kumar, MNRAS, 461, L122 (2016).

\bibitem{2018MNRAS.481.2320L} R. Luo, K. Lee, and D. R. Lorimer et al., MNRAS, 481, 2320 (2018).

\bibitem{2022MNRAS.511.1961H} T. Hashimoto, T. Goto, and B. H. Chen et al., MNRAS, 511, 1961 (2022).

\bibitem{2023SciA....9F6198Z} W. Zhu, H. Xu, and D. Zhou et al., Sci. Adv. 9, eadf6198 (2023).

\bibitem{2024MNRAS.528L.133Y} S. Yamasaki, E. Göğüş, and T. Hashimoto, MNRAS, 528, L133 (2024).

\bibitem{2020MNRAS.496.4565C} M. Caleb, B. W. Stappers, and E. D. Barr et al., MNRAS, 496, 4565 (2020).

\bibitem{2022RAA....22l4004N} J.-R. Niu, W.-W. Zhu, and B. Zhang et al., Res. Astron. Astrophys. 22, 124004 (2022).

\bibitem{2022Natur.606..873N} C.-H. Niu, K. Aggarwal, and D. Li et al., Nature, 606, 873 (2022).

\bibitem{Zhou2026} D. Zhou, P. Wang, and J. Fang et al., Sci. China-Phys. Mech. Astron. 69, 249512 (2026).

\bibitem{1991PNAS...88.2297P} S. M. Pincus, Proc. Natl. Acad. Sci. USA 88, 2297 (1991).

\bibitem{2019NatSR...912761D} A. Delgado-Bonal, Sci. Rep. 9, 12761 (2019).

\bibitem{2010cfsm.book.....C} M. Cencini, F. Cecconi, and A. Vulpiani, \textit{Chaos: From Simple Models to Complex Systems} (World Scientific, Singapore, 2010).

\bibitem{2024MNRAS.533..872S} Y. Sang and H.-N. Lin, MNRAS, 533, 872 (2024).

\bibitem{2023Sci...380..599A} R. Anna-Thomas, L. Connor, and S. Dai et al., Science, 380, 599 (2023).

\bibitem{2025arXiv250714707Z} J.-S. Zhang, T.-C. Wang, and P. Wang et al., arXiv:2507.14707 (2025).

\bibitem{2015ApJS..218...11C} A. C. Collazzi, C. Kouveliotou, and A. J. van der Horst et al., \apjs, 218, 11 (2015).

\bibitem{2020ApJ...904L..21Y} G. Younes, T. G\"uver, and C. Kouveliotou et al., \apjl, 904, L21 (2020).

\bibitem{2024ApJ...976...99S} Y.-X. Shao, P. Zhou, and X.-D. Li et al., \apj, 976, 99 (2024).

\bibitem{2024ApJS..275...39W} P. Wang, J. Li, and L. Ji et al., \apjs, 275, 39 (2024).

\bibitem{2022MNRAS.510.4049B} A. Basu, B. Shaw, and D. Antonopoulou et al., MNRAS, 510, 4049 (2022).

\bibitem{Manchester2005} R. N. Manchester, G. B. Hobbs, and A. Teoh et al., AJ, 129, 1993 (2005).

\bibitem{Espinoza2011} C. M. Espinoza, A. G. Lyne, and B. W. Stappers et al., MNRAS, 414, 1679 (2011).

\bibitem{Bath1966} M. B\aa th, Phys. Chem. Earth, 7, 115 (1966).


\bibitem{scholzel2019nonlinear} C. Sch\"olzel, Nonlinear measures for dynamical systems, Zenodo (2019).


\bibitem{Eckmann1986} J. P. Eckmann, S. O. Kamphorst, D. Ruelle, and S. Ciliberto, Phys. Rev. A 34, 4971 (1986).

\bibitem{2024ATel16420....1S} K. Shin and CHIME/FRB Collaboration, \atel, 16420, 1 (2024).

\bibitem{2024ATel16430....1U} P. A. Uttarkar, P. Kumar, and M. E. Lower et al., \atel, 16430, 1 (2024).

\bibitem{2024ATel16432....1O} O. S. Ould-Boukattine, J. W. T. Hessels, and F. Kirsten et al., \atel, 16432, 1 (2024).

\bibitem{2024ATel16434....1P} D. Pelliciari, A. Geminardi, and G. Bernardi et al., \atel, 16434, 1 (2024).

\bibitem{2024ATel16452....1K} A. Kumar, Y. Maan, and Y. Bhusare, \atel, 16452, 1 (2024).

\bibitem{2026MNRAS.545f2043C} Y.-N. Chen, Y.-K. Zhang, and Z.-G. Dai, MNRAS, 545, staf2043 (2026).

\bibitem{2025ApJ...988...62L}
J.-W. Luo, J.-R. Niu, and  W.-Y. Wang et al., \apj, 988, 62 (2025).

\bibitem{2016ApJ...829...27D} Z. G. Dai, J. S. Wang, and X. F. Wu et al., \apj, 829, 27 (2016).

\bibitem{2020ApJ...900L..21Y} Y. Yuan, A. M. Beloborodov, and A. Y. Chen et al., \apjl, 900, L21 (2020).

\bibitem{2021Univ....7...56L} Y. Lyubarsky, Universe, 7, 56 (2021).

\bibitem{2022ApJ...932L..20M} J. F. Mahlmann, A. A. Philippov, and A. Levinson et al., \apjl, 932, L20 (2022).



\end{thebibliography}
\end{document}